\documentclass[fleqn,twoside]{article}
\usepackage{espcrc2_ic}

\usepackage{graphicx}
\usepackage[figuresright]{rotating}

\intextsep=\floatsep

\newcommand{\AmS}{{\protect\the\textfont2
  A\kern-.1667em\lower.5ex\hbox{M}\kern-.125emS}}

\title{Search for neutrino point sources with IceCube 22-strings}

\author{J. L. Bazo Alba \address[DESY]{DESY, D-15738 Zeuthen, Germany}, for the IceCube Collaboration%
  \thanks{See http://www.icecube.wisc.edu for a full list of authors.}
  \\E-mail: jose.luis.bazo.alba@desy.de
}

\begin{document}


\begin{abstract}
The IceCube detector took data in its 22-string configuration in 2007-2008. This data has been analyzed to search for extraterrestrial point sources of neutrinos using several methods. Two main methods are discussed and compared here: the binned and the unbinned maximum likelihood method. The best sky-averaged sensitivity (90\% C.L.) is $E^{2}\Phi_{\nu_{\mu}}=1.3 \times 10^{-11}$ TeVcm$^{-2}$s$^{-1}$ to a generic E$^{-2}$ flux of $\nu_{\mu}$ over the energy range from 3 TeV to 3 PeV. No neutrino point sources are found from the individual directions of a pre-selected catalogue nor in a search extended to the northern sky. Limits are improved by a factor of two compared to the total statistics collected with the AMANDA-II detector and represent the best results to date.
\vspace{1pc}
\end{abstract}

\maketitle

\section{Introduction}

IceCube \cite{Ach06} is a neutrino telescope located at the South Pole. Its major goal is to discover high energy neutrinos of extraterrestrial origin. Its vertical strings, each equiped with 60 optical modules (OMs), are being deployed year by year in the ice at depths between 1.45 and 2.45 km. In the 2007/08 season 22 strings (IC-22) were operating. Results from this dataset are given here. By 2011 IceCube will be completed with 80 strings extending to 1 km$^3$ instrumented volume. 

\section{Methods and Event Selection}

Two independent analyses developed for point source searches in the TeV - PeV energy region are presented here: a binned method, an event counting approach using a fixed circular angular search bin; and an unbinned method, a likelihood approach based on the point spread function and a simple energy estimator. In a declination band, a negligible point source neutrino contribution compared to the total background is assumed, allowing an estimate of the background from scrambling data.

Muon tracks are reconstructed using likelihood algorithms, taking into account recorded hit times and charges. Neutrino event selection for both analyses is based on similar parameters: angular cuts, which partially filter out southern sky atmospheric muons ($\mu_{atm}$), and track quality parameters, which help further reject mis-reconstructed $\mu_{atm}$. These quality parameters include angular uncertainty estimates, the unscattered photon hits with a small time residual w.r.t. the Cherenkov cone, a likelihood ratio comparing a prior of a $\mu_{atm}$ track to the reconstructed one, a reduced likelihood of the directional reconstruction, etc. The main source of residual background is atmospheric neutrinos ${\nu_{\mu}}_{atm}$.

\begin{figure*}[tbp]
\centering
\includegraphics[width=51mm]{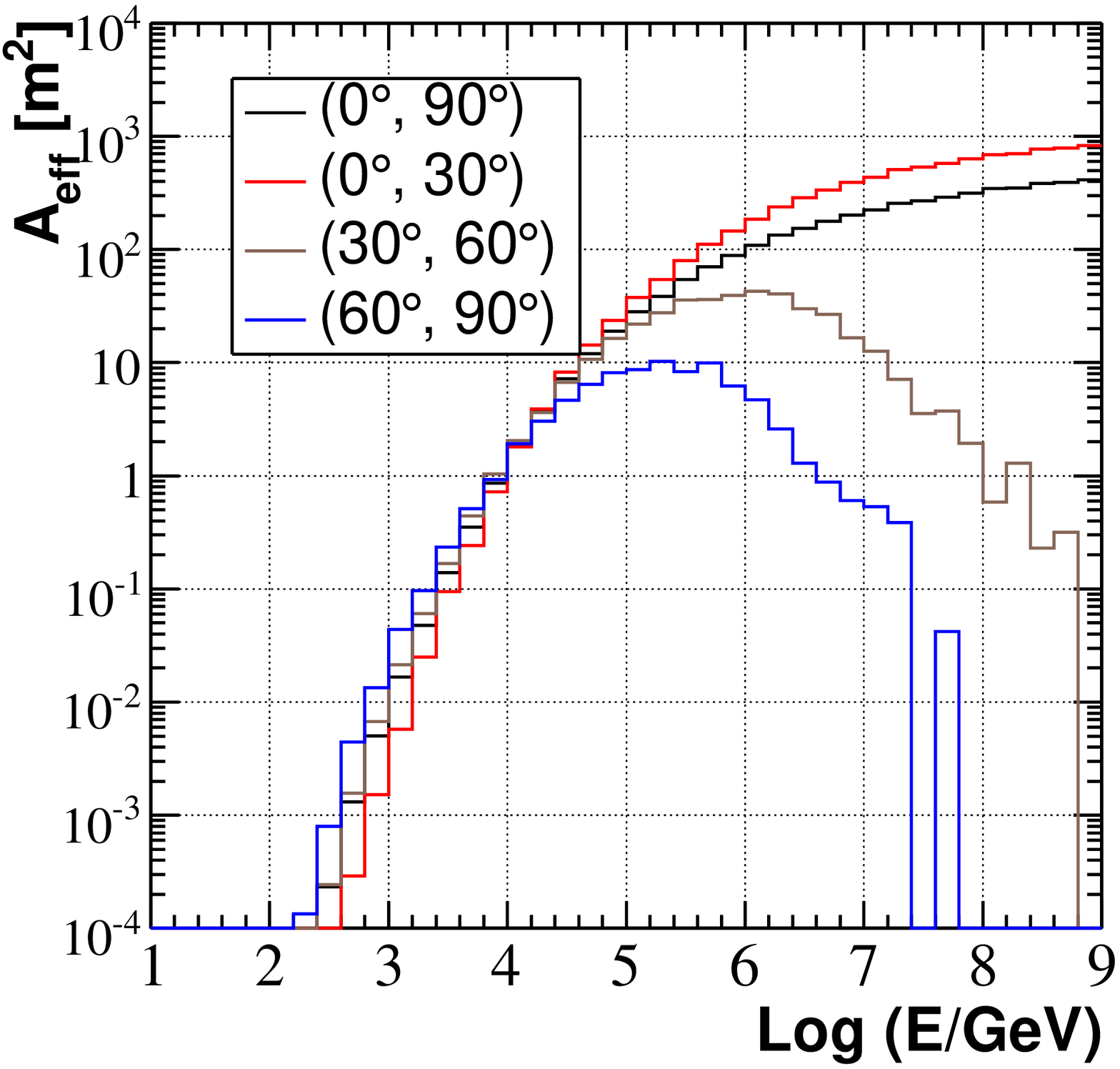} 
\includegraphics[width=81mm]{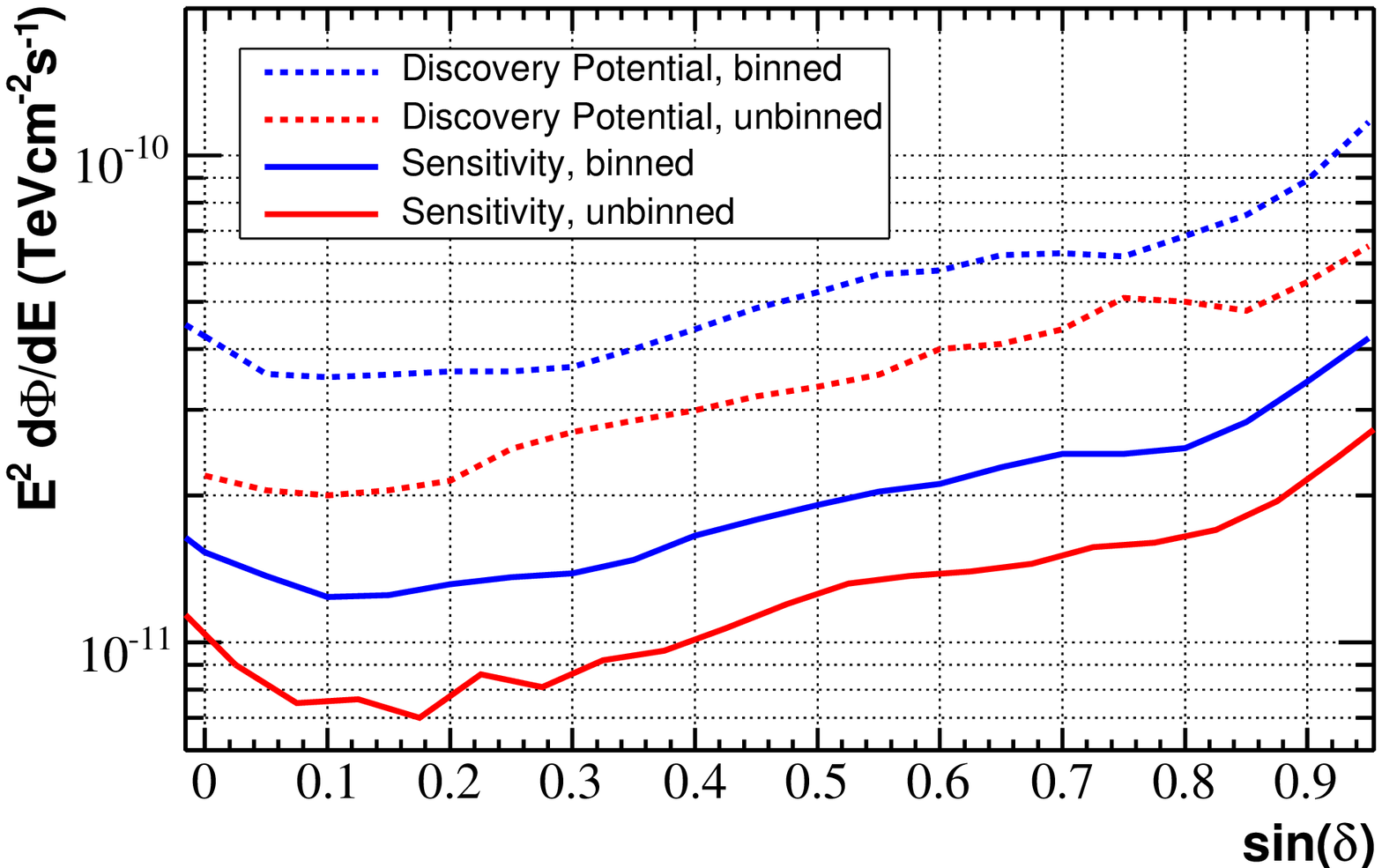}
\caption{Left: IC-22 neutrino effective area in declination ranges (binned method). Right: IC-22 sensitivity and discovery potential to point sources with $E^{-2}$ spectrum as functions of declination. }
\label{fig:area_sen}
\end{figure*}

\subsection{Binned search}
The binned method \cite{Acker07} distinguishes a localized excess of signal from a uniform background (${\nu_{\mu}}_{atm}$ and misreconstructed $\mu_{atm}$) using a circular angular search bin. The sensitivity, in a given declination band, is calculated by comparing the number of simulated signal events reconstructed inside the search bin, $n_{sig}$, to the average number of background events from data, using: 
\begin{equation}
  \Phi=\Phi_0 \frac{FC_{\mu_{90}}(n_{bg})}{n_{sig}},
  \label{sen_binned}
\end{equation}
where $\Phi_0$ is the signal flux normalization, $FC_{\mu_{90}}$, given by Poisson statistics in the Feldman \& Cousins approach \cite{FC98} represents the average event upper limit (90\% C.L.) for the expected background in the bin, $n_{bg}$, and no true signal.  

Cuts on track quality parameters are optimized considering the variation in detection and reconstruction efficiencies with declination. In addition, the search bin radius depends on declination (mean value = 2.1$^{\circ}$). Final cuts were chosen so as to optimize sensitivity for a compromise between $E^{-2}$ and $E^{-3}$ neutrino energy spectra. 

\begin{figure*}[!tbp]
\centering
\includegraphics[width=116mm]{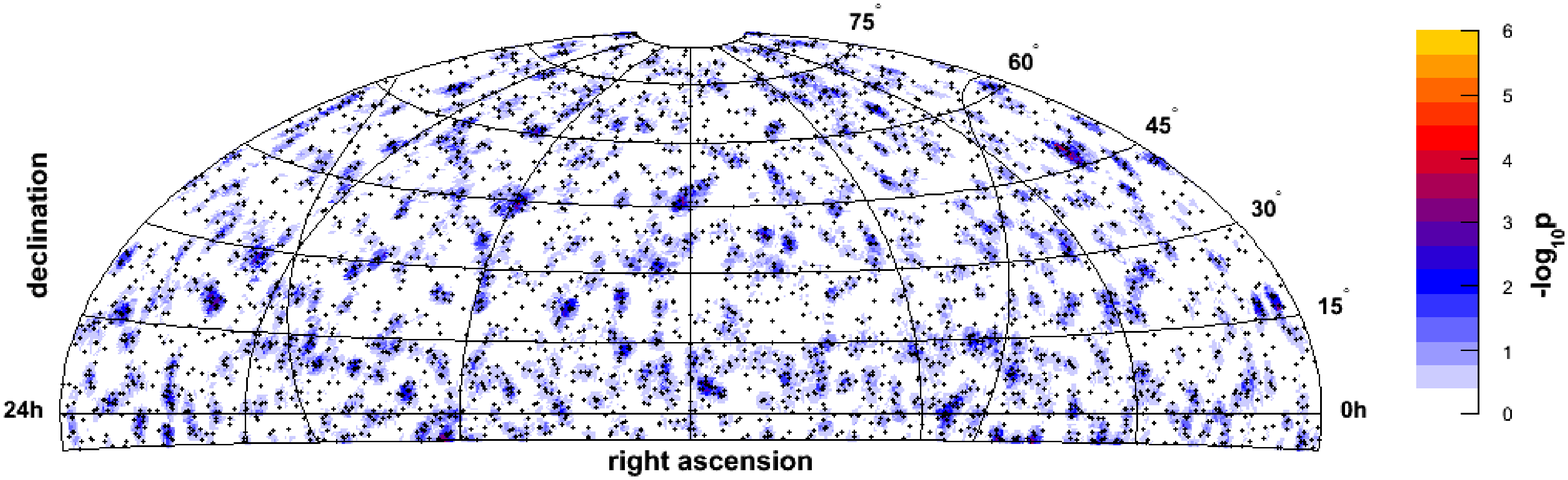}
\includegraphics[width=116mm]{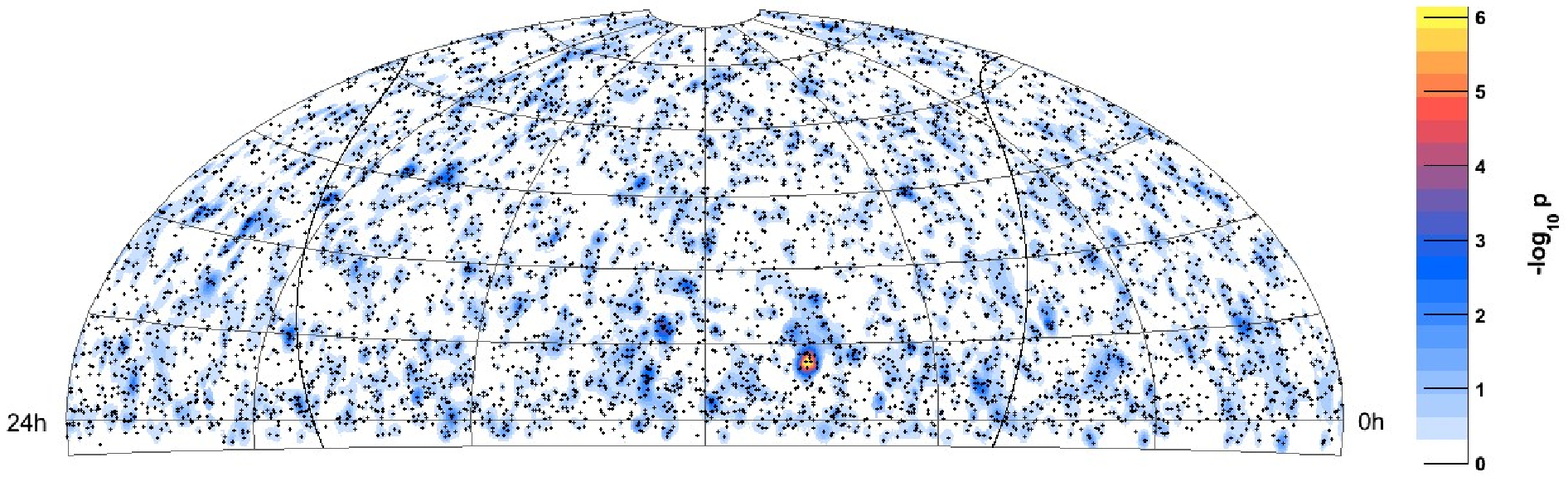}
\caption{IC-22 skymap with pre-trial p-values (in colors) and event locations (dots). Top: binned method. Bottom: unbinned method. Each method uses a different direction reconstruction technique.}
\label{fig:skymap}
\end{figure*}

\subsection{Unbinned maximum likelihood search}
The unbinned maximum likelihood method \cite{Braun08} constructs a likelihood function, ${\cal L}$, which depends on the signal probability density function (pdf) and background pdf, for a given source location, $x_{s}$, and total number of data events $n_{tot}$:

\begin{equation}
  {\cal L}=\prod_{i=1}^{n_{tot}} \left ( \frac{n_{s}}{n_{tot}}S_{i}(x_i,x_{s},E_i,\gamma)+ \frac{n_{bg}}{n_{tot}} B_i(x_i,E_i) \right ).
  \label{llh_unbinned}
\end{equation}

The signal pdf, $S_{i}$, is characterized by each event's reconstructed direction, $x_{i}$, angular uncertainty around $x_{s}$, energy estimation, $E_i$, i.e. the number of OMs hit, and $\gamma$, the assumed energy spectrum index ($E^{-\gamma}$). The background pdf,  $B_i$, is determined from the declination distribution and energy estimators of the data sample. The number of signal events, $n_s$, is found by maximizing the likelihood ratio of the background plus signal hypothesis against the background-only case.
 
This method does not need tight cuts, as in the binned analysis, because the likelihood incorporates energy information that can separate signal from background. A harder event selection does not further improve the final sensitivity.

\section{Data Samples}

The declination range from -5$^{\circ}$ to 85$^{\circ}$ has been scanned in the TeV-PeV energy region using IC-22 data with a lifetime of 276 days. Each analysis has followed its own event selection criteria, arriving at a final sample of 5114 (2956) events for the unbinned (binned) method. Since the binned method needs more efficient cuts to get a better sensitivity, the final sample is smaller than the unbinned method, which contains more low energy ${\nu_{\mu}}_{atm}$. The $\nu_{\mu_{atm}}$ content of the samples is greater than 90\%, where the contamination comes from misreconstructed down-going muons, mostly time-coincident multiple events. 

From simulation, a sky-averaged median angular resolution of $1.4{^\circ}$ is estimated for signal neutrinos with E$^{-2}$ spectrum. The final effective area for a $\nu_{\mu}+\bar{\nu}_{\mu}$ flux is presented in Fig. \ref{fig:area_sen} and is similar for both analyses. The effect of the absorption in the earth, due to increased cross section with energy, can be seen for events at large declination.

\section{Results}

Sensitivities and discovery potentials\footnote[1]{Smallest flux detectable at $5\sigma$ in 50\% of trials.}, as functions of declination, are compared in Fig.~\ref{fig:area_sen}. The sky-average senstitivity at 90\% C.L. to a generic E$^{-2}$ flux of $\nu_{\mu}$ is 1.3 (2.0) $\times 10^{-11}$ TeV$^{-1}$cm$^{-2}$s$^{-1}$ (E/TeV)$^{-2}$ for the unbinned (binned) method. The enhancement, without including systematic uncertainties, gained by using the unbinned method is 35\%. The unbinned sensitivity shows a factor of 2 improvement over that of the total statistics collected by AMANDA-II \cite{AMANDA7yr} and represent the best limits to date.

A pre-defined list of 28 sources (galactic and extragalactic) gave no significant excess over background. The major deviation was found for 1ES 1959+650 (Crab Nebula) with a post-trial p-value of 66\% (91\%) by the unbinned (binned) analysis.

The results from the all-sky search for both analyses are shown in Fig.~\ref{fig:skymap}. The maximum upward deviation from the background hypothesis is located at r.a. = $153^{\circ}$ ($183.5^{\circ}$) , dec = $11^{\circ}$ ($47^{\circ}$), with a post-trial p-value of 0.67\% (67.6\%) for the unbinned (binned) analysis. The final corrected p-value for the unbinned analysis is 1.3\%. This represents the overall outcome of IC-22, as it was decided a priori that the final result would be based on the more sensitive unbinned analysis. The unbinned-spot, to be verified with IC-40 data, was further investigated (e.g. time-dependent analysis) resulting in no significant event cluster in time (no contributing events are closer together than 10 days).

\section{Summary \& Outlook}

The results of both methods are consistent with background fluctuations. No neutrino point sources are found from the individual directions of the pre-selected catalogue nor in the extended northern sky search. The limits set with IC-22 are the best to date. Systematic uncertainties will be included in the final results. 

IceCube is half completed with 40 strings taking data since April 08. A better sensitivity is expected given the bigger effective area and better angular resolution. In the coming 2008-09 season, 16 more strings are planned to be deployed.

\end{document}